# STUDYING ACADEMIC INDICATORS WITHIN VIRTUAL LEARNING ENVIRONMENT USING EDUCATIONAL DATA MINING

Eng. Eid Aldikanji[1] and Dr. Khalil Ajami[2]

[1]Master Web Science, Syrian Virtual University, Damascus, Syria
[2]Syrian Virtual University Vice President, Bachelor of Information Technology Program Director, Damascus, Syria

## ABSTRACT

*The rapid developments in information and communication technologies taking place recent years, make it easy for people to pursue their education distantly. The development of new technologies helped to open spatial and temporal boundaries of learning toward virtual learning, and helped to transform education process from its classical form of courses within classrooms to a new virtual form within virtual environments; Consequently, lessons and lectures are delivered using information and communication technologies tools, and student's attendance is virtually performed via Internet. Moreover, the education process in its new form becomes a supervised process, rather than a fully controlled process since virtual learning changed the education process pattern represented by the triangle (student, teacher and content) by increasing the importance of both "student" and "content" factors and transforming the main task of the "teacher" from "Teaching" to "Tutoring". Consequently, many questions are raised concerning students' performance and concerning the adequacy of virtual learning process. These questions are related to the need of accreditation for virtual learning and virtual universities.*

*Our work aims to use Educational Data Mining (EDM) in order to study academic indicators concerning a representative sample of students in a virtual learning environment within Syrian Virtual University – SVU (The students who are following Bachelor of Information Technology Diploma –BIT). Our main goal is to discover the main factors influencing students' academic trajectory and students' academic evolution within such environment.*

*Our results indicate strong correlation -in this virtual learning environment- between student average and some factors like: student's English level (despite the fact that Arabic language is the teaching language), student's age, student's gender, student's over-stay and student's place of residence (inside /outside Syria). Our results indicate also a need to modify the academic trajectory of students by changing the prerequisites of few courses delivered as a part of BIT diploma like Advanced DBA II, Data Security. In this research, the results also highlight the effect of the Syrian Crisis on students. Finally, we've suggested some future recommendations based on our observations and results to develop the current information system in SVU in order to help us to deduce some indicators more easily*.

## KEYWORDS

*Educational Data Mining; Classification; Naïve Bayes; Decision Trees; Regression Models; Syrian Virtual University, Virtual Learning Environment*

## 1. INTRODUCTION

The development of new technologies helped to transform education process from its classical form of courses within classrooms to a new virtual form within virtual environments; Consequently, lessons and lectures are delivered using information and communication technologies tools, and students' attendance is virtually performed via Internet. Moreover, the education process in its new form becomes a supervised process, rather than a fully controlled process since virtual learning changed the education process pattern represented by the triangle (student, teacher and content) by increasing the importance of both "student" and "content" factors and transforming the main task of the "teacher" from "Teaching" to "Tutoring". Consequently, many questions are raised concerning students' performance and concerning the adequacy of virtual learning process. These questions are related to the need of accreditation for virtual learning and virtual universities.

Otherwise, Syrian Virtual University (SVU) is a public academic institution established in 2002 in response to developments in e-learning. Its objective is to help develop human resources in various disciplines in order to meet the needs of economic and social developments and market needs and to keep abreast with the requirements of a knowledge-based economy, especially in areas like Information and Management Systems, E-Marketing, IT and Internet Technologies. SVU is constantly developing its infrastructure and expanding its exam centers to accommodate the rapid increase in demand on its services, since the number of SVU students actually reached 30 thousand students.

Regionally, The SVU is also contributing to the development of legislations and regulations governing virtual education besides establishing the criteria to be met by institutions delivering education in this mode which will facilitate the work of other virtual educational institutions and will improve the employability of SVU graduates in the Arab labor market. In this context, it should be emphasized that the knowledge and skills demonstrated by graduates from virtual learning environments is no less than that demonstrated by graduates from the traditional educational systems

Due to this variety, SVU administration needs to know more about students' performance and about students' metrics affecting their studies in order to improve SVU services, increase students grades [1], enhance educational outcomes, and explain some educational phenomena [2]. This also helps to obtain new knowledge and offers suggestions for academic planners [3] helping them to predict students' success/failure and drop-off [4]. Such an enhancement improves the education process for SVU students and guarantee worldwide academic accreditation for SVU and for virtual learning in general.

Technically speaking, the study of such metrics is based on SVU Students Information System SVU-SIS (www.svuonline.org) using Educational Data Mining (EDM) which is the application of Data Mining (DM) techniques to educational data [6]. In fact, we perform our work on a representative sample of students constituted of BIT students (Bachelor in Information Technology Program). We study the evolution of such a sample from 2006. Data extracted include information about students like name, mother, father, phone, courses, detailed graded in each term for each course, exams information. We used only (student_id) field as a key item, removing every personal details for privacy reasons [7].

This paper is organized as follows: Section 2 describes literature studies of the previous works. Section 3 describes an introduction to Educational Data Mining. Section 4 describes the research methodology used and the students' data in our study, in addition to describing the process steps. Section 5 describes applying regression, classification and decision trees on the preprocessed Data in order to deliver our results. Section 6 describes recommendations to develop the current SVU Student Information System, final conclusion is on section 7.

## 1.1 Abbreviations and Acronym

Table 1 Abbreviations

| Abbreviations | Description | Abbreviations | Description |
|---|---|---|---|
| **EDM** | Educational Data Mining | **Min-Red** | Minimum Redundancy |
| **DM** | Data Mining | **Max-Rel** | Maximum Relevance |
| **SVU** | Syrian Virtual University | **KNN** | K-Nearest Neighbor |
| **SVUIS** | SVU Information System | **NN** | Neural Network |
| **BIT** | Bachelor of Information Technology in SVU | **CS** | Computer Science |
| **NB** | Naïve Bayes | **RB** | Rule-Based |
| **DT** | Decision Trees | **ID3** | Iterative Dichotomiser 3 |
| **GA** | Genetic Algorithm | **LR** | Linear Regression |
| **SVM** | Support Victor Machine | **LoR** | Logistic Regression |
| **IG** | Information Gain | | |

## 2. LITERATURE STUDIES

Studies about EDM and LA have b very common in the last two decades and are used by a lot of institutes and universities around the world. Whereas, here in Syria, this is the first research in this field, and is being applied as a part of Master Web Science Thesis. In this section we're going mention some of these studies:

Dr. P. Nithya et al. explained it in their survey paper published 2016 in IJARCET [8], Karan et al. published a review toward knowledge engineering in EDM [9], Merceron published a paper concludes challenges on the way of educational analysis, methods, tasks and current trends [4]. We also never forget the God Fathers of EDM C. Romero, S. Ventura, R S.J.d Baker in their papers [10] [6] [11] [12] [13] [14] [15].

W. Punlumjeak and N. Rachbure [16] proposed a comparison of four feature selection methods GAs, SVM, IG, Min-Red and Max-Rel four supervised classifiers: NB, DT,KNN , and NN. They found that Min-Red and Max-Rel is the most appropriate way with accuracy 99.12% using KNN. A. A. Aziz, N. Hafieza, and I. Ahmad [17] from UniSZA proposed a framework for predicting students' academic performance of 1st year bachelor students in CS course between July 2006/07 and July 2013/14 using NB, DT, RB Classification and found that RB is the best model with accuracy 71.3%.

L. Sibanda, C. G. Iwu, and O. H. Benedict [18] studied the factors influencing student success/failure in educational process, and they found that regular attendance at lecture is the most influencing factor on student success, this result supports L. P. Steenkamp, R. S. J. d. Baker, and B. L. Frick study [19]. In failure factors, they found the noisy lecturing and not finishing or doing assignments is the most influencing one's.

K. Adhatrao, A. Gaykar, A. Dhawan, R. Jha, and V. Honrao [20] predicted the general and individual performance of freshly admitted students in future examinations by applying ID3 and C4.5 on the historical students' information, exams, marks, grades of classes X and XII, entrance examinations and results in first year. They scored accuracy equals 75.275%.

International Journal of Data Mining & Knowledge Management Process (IJDKP) Vol.6, No.6, November 2016

## 3. INDRODUCTION TO EDM

EDM is an interesting research area which extracts useful, previously unknown patterns from educational databases using DM. The obtained knowledge then can be used for better understanding, improving educational performance and assessment of the student learning process, in addition to discovering how people learn, predicting learning, and understanding real learning behavior. By achieving these goals, EDM can be used to design better and smarter learning technology and to better inform learners and educators [6][21][11].

EDM uses DM techniques and methods, and usually DM process consists of two phases training and testing. In the first, the model is built using training data set where each tuple is being analyzed depending in the algorithm, while in the latter the constructed model is used to assign label to an unlabeled training data set [22]. Before applying DM we need to pre-process data to make it suitable for applying algorithms. This step is considered the most important step, and the quality of results depends on it. Usually, this step take between 60-90% of manual time [23][10].
In this paper we use classification techniques like NB, LR, LoR, DT [ID3, C4.5] with different attributes selecting measures to predict students indicators, correlations between courses, student's statistics after Syrian Crisis and other presented factors.

## 4. PROCCESSING BIT DATA AND RESEARCH METHODOLOGY

Selecting the most appropriate DM algorithm depends on the deep understanding to the academic environment, the expected outcomes and the pre-analyzing of the educational data. We've extracted the data for this research from SVUIS databases for graduated BIT students between Fall 2003 and 2014, in total 8344 [1691 graduated] students, 146967 different transactions on their profiles in 42 course.

As any information system, data was stored in different types and formats like databases, Excel files, flat files and XML files. These formats are not usually ready to be applied directly to DM algorithms, so they need preparation first, and this phase explained with all the steps for students' performance measurement in (Figure 1)

The original form of data contains a lot of features and attributes describing students and courses, so we have worked with domain experts in SVU to use some feature selection techniques to minimize redundancy and maximize relevance feature subset while retaining a high accuracy without losing any important information about students. The effective process of feature selection can play as an important role by improving learning performance, lowering computational complexity, and building better generalizable models [16][24].

After features are selected, we have converted and consolidate data by migrating it into a MySQL DB to be prepared before applying different DM algorithms. Some of steps we used is to transform between different types, clean and integrate data, remove redundant records for dataset in general and for each algorithm.

After this phase (preparation phase) data is now ready, and the final table of features presented in(Table 2 describes each features and its value type. The detailed description about them in (Table 3)

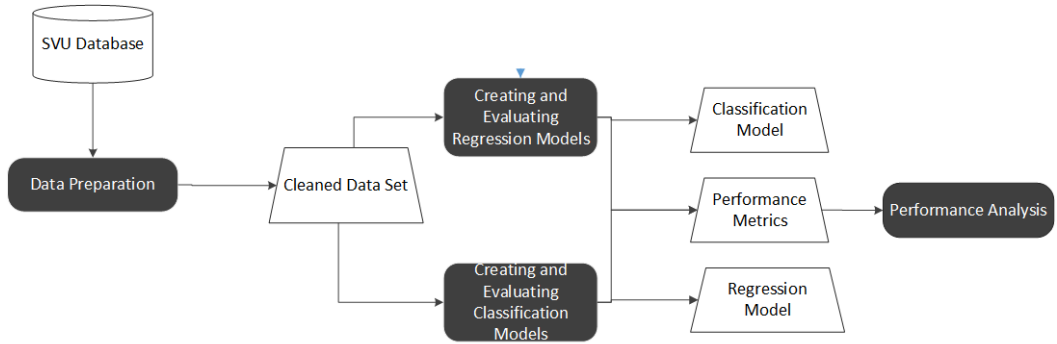

Figure 1. Students' performance mesaurment steps

## 5. APPLYING EDM TO BIT DATA

In this section we will explain the results acquired from the applying DM algorithms like regression [Linear and Logistic] and classification [Naïve Bayes, ID3 and C4.5 decision trees].
We used RapidMiner to apply the algorithms and get the results. For each algorithm, 10-fold cross validation with 0.1:0.9 and 0.2:0.8 was used to aquire maximum efficiency for each algorithm, and to decrease error rate. After that, we have applied the same model to unlabeled data to check the accuracy and other performance metrics like Kappa, error rate, correlation. In the folowing, we descibe the results and observations.

### 5.1 General Results

#### 5.1.1 Relation between student's average and the country of residence

Students with place of residence outside Syria have scored a better average than students inside. This indicator is one of the important indicators need to be studied in details to find the reasons and results affecting on students, considering the social environment in Syria.

#### 5.1.2 Relation between student's average and English level

Students with "good" English level only have less average than students with excellent or TOEFL/IELTS certificate as shown in (Figure 2).This indicator also found when studying the effect of English level on several courses in BIT like Operating Systems [ITI310/320], Advanced SQL Programming [ITD310] and this indicator became more clear in advanced courses like Data Security [ITI420]. Through this results we suggest to increase English level for the new enrollment students to increase their efficiency.

#### 5.1.3 Relation between student's average and age

We've found that students with "up_normal" age [more than the average age] have higher average that students with "normal" age.

Table 2 Final attributes and features for BIT students

| W | Data | Attribute | Data | Attribute | Data |
|---|---|---|---|---|---|
| **student_id** | 00000 | **cis101** | ## | **ita330** | ## |
| **term_name** | Fall /Spring | **itc110** | ## | **itad430** | ## |
| **gender** | Male/female | **itc200** | ## | **itd410** | ## |
| **age_text** | Normal/up_normal | **itc210** | ## | **itd420** | ## |
| **mention** | Good/V.Good/Excellent | **itc220** | ## | **iti410** | ## |
| **country** | Syria/Out_Syria | **itc230** | ## | **iti420** | ## |
| **specialization_name** | BI/BC/BF/BH | **itd300** | ## | **iti300** | ## |
| **telecenter_name** | NAME | **itd310** | ## | **iti310** | ## |
| **discount** | no_discount/discount | **ita310** | ## | **iti320** | ## |
| **Eng** | Good/v.g/excellent/t_i_cert | **ita320** | ## | **bat_pt** | ## |
| **stay_text** | Normal/up_normal | **unv101** | ## | **grad_itp390** | ## |
| | | **Average** | ##.## | | |

Table 3 Attributes description

| Attribute | Description | Attribute | Description |
|---|---|---|---|
| **student_id** | Student identification number | **ita330** | Business Process Modeling |
| **term_name** | Registration term | **itad430** | Mobile Application |
| **gender** | Student gender | **itd410** | MS-SQL Development and Administration |
| **age_text** | Student age category | **itd420** | Advanced Database Administration II |
| **mention** | Student graduation mention | **iti410** | Data Security |
| **country** | Student country | **iti420** | Data Networks |
| **specialization_name** | Student specialization | **iti300** | Network Operating Systems |
| **telecenter_name** | Exams telecenter | **iti310** | Windows Platform |
| **discount** | Student study with discount | **iti320** | Linux Platform |
| **Eng** | English level | **bat_pt** | BAT PT |
| **stay_text** | Student Sojourn Period | **grad_itp390** | Internship |
| **cis101** | ICDL | **itd310** | Advanced SQL Programming |
| **itc110** | Introduction to Operating Systems | **ita310** | Object Oriented Programming and Design |
| **itc200** | Introduction to Networks | **ita320** | Web Application Design and Development |
| **itc210** | Introduction to Data Modeling | **unv101** | Introduction to On-Line Education |
| **itc220** | Web Development and E-Commerce | **Average** | Student Average |
| **itc230** | Introduction to Programming | **itd300** | Data Base Architecture and Design |
| **itd415** | Advanced Data Base Administration I | | |

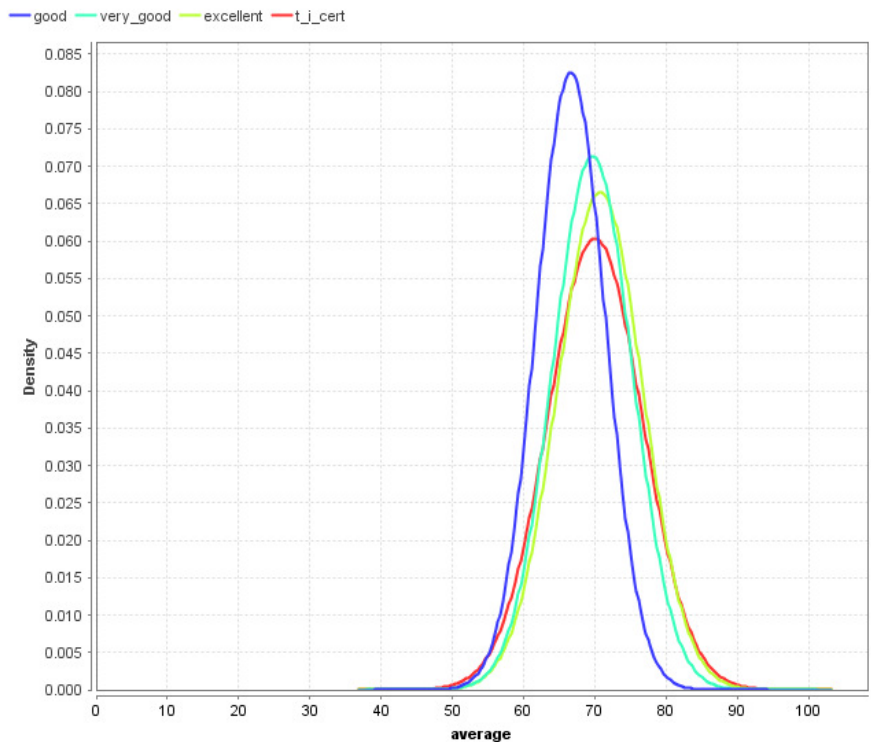

Figure 2 Relation between AVG and ENG

### 5.1.4 Relation between student's average and gender

We have found a very strong relation between the high average for females compared to males. This indicator can be interpreted as in Syria; males spend a more time studying to avoid the military service and the repeated failing in courses decrease the average and performance of the students. We have also found the increment of low-averages after the Syrian Crisis specially after 2012.

### 5.1.5 Relation between Sojourn Period (stay at university) and student's average

Here we've discovered the negative influence of staying more at university and the average, in this discovery the over time that students stays at university affecting negatively on their performance and final average. And also, the students with normal stay rate are better that others. There is a little relation between the over-stay and gender male, as explained in last section [**Error! Reference source not found.**]

### 5.1.6 Increment of students after Syrian Crisis

One of the most significant indicators in this research is we found the high increment of student after the Syrian Crisis since 2011, and the number double after year 2013. Such observations maybe because young people escaping from military service and war into studying, specially e-learning and virtual classes (Class 5)

35

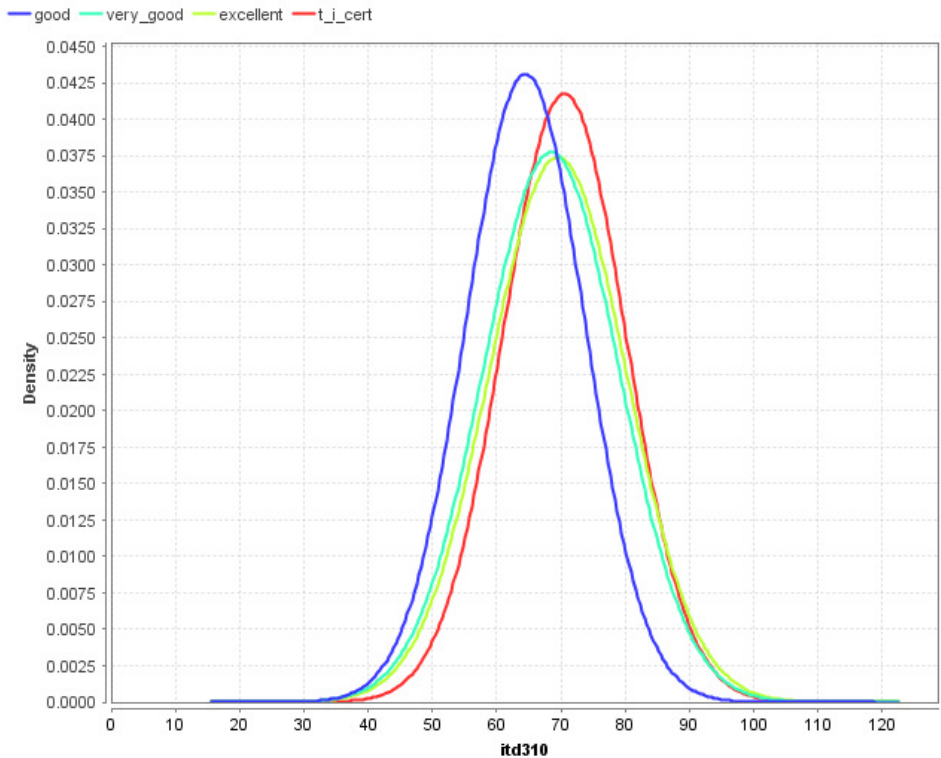

Figure 3 Relation between ITD310 and ENG

## 5.2 Applying NB

We applied NB to predict mention with three classes (3C – good, very-good, excellent) accuracy was 56.76% for BH students, for BF-BC students the model accuracy was 73%. This increment is because the increased number of students in these two specialization. When we add ITCXXX courses, the model accuracy became 78% with about 5% improvement.

When we used two classes (2C – good and excellent), we noticed the accuracy for BH increased to 87.44%, for BF-BC accuracy 87.43% and with ITCXXX became 93.31%. For all specializations combined accuracy became 87.98%. even with this high accuracy, the model is not very good for predicting excellent class because the low number of excellent classes.
Applying NB to predict the sojourn period produce a model with 79.97% accuracy.

We discovered using NB a very good influence between the up normal age and the average, we also found that females' average is higher than males. The latter observation maybe because the military service and its influence on Syrian males, so they fail themselves in more courses to stay at university instead of going to war.

International Journal of Data Mining & Knowledge Management Process (IJDKP) Vol.6, No.6, November 2016

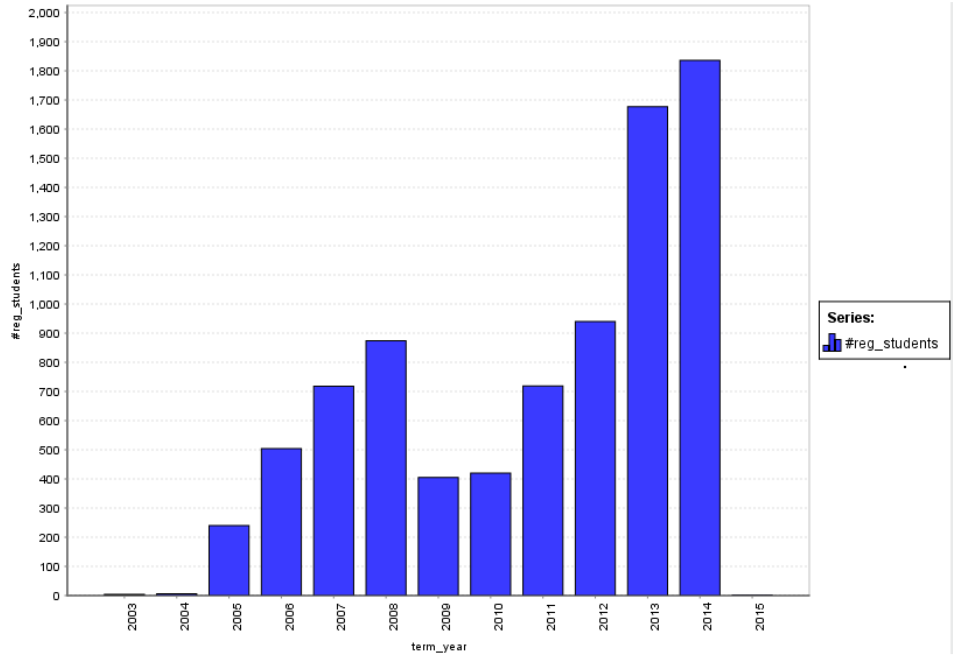

Figure 4 Increment of students number after Syrian Crisis

## 5.3 Applying LR

The LR model predicted students' average with error rate less than 5.7%, and found that (excellent and very_good) English students have positive influence on average. Student registered in Fall have negative influence on average, in contrast Spring has positive influence on average.

## 5.4 Applying LoR

LoR needs a special type of attributes, only numeric ones with binary label class (good – excellent). So we transformed the attributes and built the model. Accuracy in this situation was 78.76% for predicting mention, and 65.99% for predicting sojourn period.

## 5.5 Applying DT

Decision Trees usually are the most used techniques in EDM, because of the clarity and easily to interpret tree.

### 5.5.1 ID3

This algorithm requires only categorical attributes, so we applied transformations on the attributes and built the model. The accuracy to 3C was 64.87%, 3C with ITCXXX was 75.60%, (Figure 5) display a part of ID3 tree. When we build 2C model, accuracy increased to 87.64%, with ITCXXX decreased to 84.62% but the model became more accurate while predicting Excellent class. For sojourn period accuracy was 78.51%.

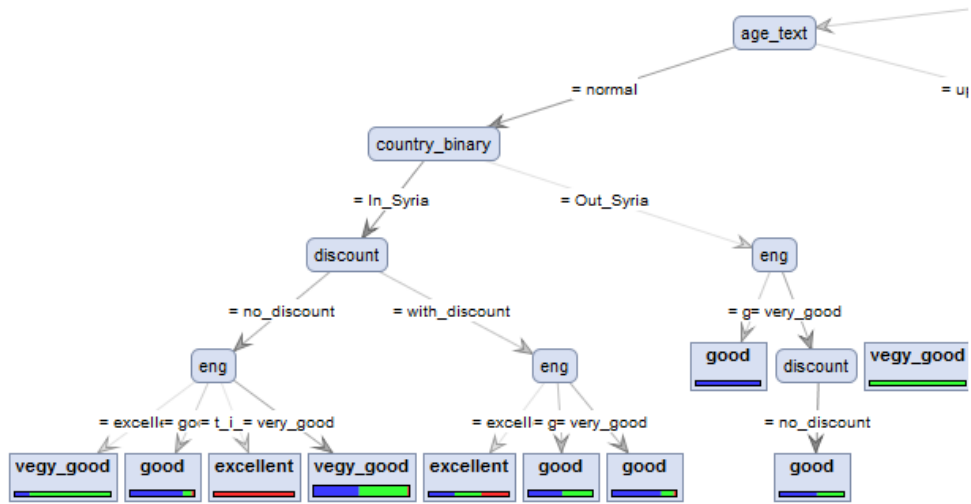

Figure 5 Part of ID3 Tree - 3C

### 5.5.2 C4.5

The built model for 3C achieved 64.95% accuracy, with ITCXXX accuracy increased to 74.67%. For 2C accuracy was 87.96% and better for excellent class. When we added ITCXXX accuracy decreased to 74.67%. For predicting sojourn period C4.5 achieved 78.92.

## 5.6 Studing correlations between courses and their prerequisite

In this scenario we have tried to find correlations between couple of courses and their prerequisite to try some suggestions that helps in increasing students' performance.

### 5.6.1 Advanced Database Administration II(ITD420) relation with its pre-requisite and other courses

We've found a strong relation between "Advanced Database Administration II" and [ Advanced Data Base Administration I – ITD415], but Advanced SQL Programming (ITD310) the direct prerequisite for ITD415 is less correlated than MS-SQL Development [ITD410] and Data Base Architecture and Design [ITD300]. We suggest here to alter the prerequisite for Advanced DBA by adding MS-SQL Development as required, the same for DB Architecture and Design.
We also have found a very strong relation with English lever, so we suggest to increase English level for the advanced DBA I & II.

### 5.6.2 Data Security ITI410 pre-requisite

We have found that is this study that Data Security is so related with Operating System courses (Windows / Linux under code ITI310/320) more that the direct prerequisite "Network Operating Systems" [ITI300]. This is a direct indicator that we have to teach students Windows and Linux before Data Security course. We have also found a strong correlation with English.

### 5.6.3 Data Networks ITI420

For this course, we found a strong correlation between the Data Networks and Operating Systems courses [Windows and Linux] more than the direct prerequisite Network Operating System, we suggest here like Data Security course, to add OS Win/Linux ITI310/320 to prerequisites.

## 6. FUTURE RECOMMENDATIONS

The current information of students from this virtual environment helps to find very helpful indicators about the students, their performance and the problems may they have it. There are also a lot of attributes that helps to extend the prediction of students' performance, and these attributes need to be studied very carefully with domain experts like teacher, student affair staff and other persons related directly or indirectly with SVUIS. This needs a cumulative experience increased with the incremental understanding to the Syrian society, in the following we mention some attributes we found it important to be recorded for future recommendations

  i. **Recording more and more student's data**

   A lot of studies found strong relations between student's grades in high school and university because of the cumulative experience built during student's study, in addition to information like the school [urban, ruler], social information about the students.

  ii. **Recording living conditions**

   Some studies found the importance of living conditions influence on student performance, we maybe try to do some of these suggestions.

  iii. **Recording attributes about exams directly**

   During this research, we unable to find some direct information like:
   How many times the student re-take exam?
   How many times the student failed in a specific term?
   What is the longest time a student has re-take some courses?
   And more questions important to find the dropout indicators for students.

  iv. **Recording more data about teachers and their courses**

   We use such information to highlight the problems in course, for example if all the students have a low-average on a certain course this raises a problem-flag. Maybe the problem in course, its content, or maybe a teacher. This help us for updating courses to increase student productivity and performance

  v. **Recording students interaction and engagment with the system**

   Students' engagements with the system is one of the most significant indicators especially in a virtual learning environment. This interactions during the class, assignments and exams provide important information about student's behavior's. For example, the times of delayed assignments indicates a lazy student with bad performance, the attempts to play with the system indicates if a student is trying to cheat or not.

International Journal of Data Mining & Knowledge Management Process (IJDKP) Vol.6, No.6, November 2016

## 7. CONCLUSION

The aim of this work is to study the academic indicators and factors within virtual learning environment influencing students' academic trajectory and students' academic evolution in general and its application on Syrian Virtual University students in BIT diploma program. Another important purpose is to highlight the current information system in SVU to keep up with the accelerated developments in information systems, try making more precise statistic and predictions using EDM to solve the current inconsistencies.

One of the most significant indicators found in this study is the influence of English language level on student's average and marks in several courses (despite the fact that Arabic language is the teaching language), the influence of students age, gender, over-stay at university and the country of residency on his/her average and performance. Our results also highlight the influence of the Syrian Crisis on the increment of student's number and their over-stay at university. We have also suggested changes in the prerequisites for some courses like Data Security, Data Networks and Advanced DBA II.

In addition, we have applied classification and regression algorithms to build a model able to predict students average mention. We have acquired the best accuracy for three classes prediction 3C [good , very_good, excellent] using NB at 68.12%. For two classes 2C [Good, Excellent] C4.5 was the best with 87.96% accuracy.

In prediction student's average using LR we've predicted the average with +/- 5.237 marks max.
Our results in predicting student's stay period [normal – up_normal], we have acquired the best accuracy using NB at 79.97%. For the complete recap to all algorithms and their performance metrics, described in details in (Table 4).

Following-up the point we are now in is so important. This work can be extended to students from different programs in SVU. It can also be integrated with current SVU student information system in order to give direct insights about students' performance, and suggestions that make the system more intelligence, more elegant.

## ACKNOWLEDGMENT

I sincerely like to thank everyone who has helped in this research, first of all Dr. Khalil Ajami my teacher and thesis supervisor for his kindness, help and precious time spent with me to finish this research, Master Web Science doctors, teachers and all SVU staff and members.
I would also like to thank my family; mother, father, brothers and my beloved fiancée for their great support.Table 4 Final recap for all algorithms and performance metrics

| Algorithm | Performance Measures | | | | |
|---|---|---|---|---|---|
| C1 - Predict Mention | Accuracy | KAPPA | R | $R^2$ | RMSE |
| NB-3C:BI | 70.43% | 0.325 | 0.370 | 0.147 | 0.474 |
| NB-3C:BH | 56.67% | -0.050 | 0.122 | 0.075 | 0.536 |
| NB-3C:BF-BC | 73.69% | 0.180 | 0.402 | 0.211 | 0.456 |
| NB-3C:BF-BC ITCXXX | 78.35% | 0.463 | 0.574 | 0.361 | 0.390 |
| NB-3C:ALL | 68.12% | 0.164 | 0.242 | 0.071 | 0.489 |
| NB-2C:BI | 87.44% | 0.165 | 0.206 | 0.042 | 0.309 |
| NB-2C:BF-BC | 87.43% | 0.177 | 0.211 | 0.122 | 0.304 |
| NB-2C:BF-BC ITCXXX | 93.31% | 0.628 | 0.648 | 0.552 | 0.221 |

40

| | | | | | |
|---|---|---|---|---|---|
| **NB-2C:ALL** | 87.89% | 0.021 | 0.053 | 0.014 | 0.321 |
| **LoR-2C:ALL** | 78.76% | 0.168 | 0.176 | 0.041 | 0.318 |
| **ID3-3C:BF-BC ITCXXX** | 75.60% | 0.311 | 0.425 | 0.257 | 0.469 |
| **ID3-3C:ALL** | 64.87% | 0.112 | 0.143 | 0.026 | 0.510 |
| **ID3-2C:BF-BC ITCXXX** | 84.62% | 0.218 | 0.284 | 0.168 | 0.383 |
| **ID3-2C:ALL** | 87.64% | 0.077 | 0.131 | 0.034 | 0.331 |
| **C4.5-3C:BF-BC ITCXXX** | 74.67% | 0.292 | 0.415 | 0.255 | 0.481 |
| **C4.5-3C:ALL** | 64.95% | 0.089 | 0.132 | 0.022 | 0.499 |
| **C4.5-2C:BF-BC ITCXXX** | 90.90% | 0.336 | 0.520 | 0.270 | 0.255 |
| **C4.5-2C:ALL** | 87.96% | 0.074 | 0.219 | 0.048 | 0.305 |
| | | | | | |
| **C2 – Predict AVG** | **Prediction AVG** | | **R** | **R$^2$** | **RMSE** |
| **Linear Regression** | 67.949 +/- 5.237 | | 0.291 | 0.085 | 5.040 |
| **C3 - Predict Stay** | **Accuracy** | **KAPPA** | **R** | **R$^2$** | **RMSE** |
| **NB:ALL** | 79.97% | 0.077 | 0.370 | 0.147 | 0.474 |
| **LoR:BH** | 65.99% | 0.198 | 0.217 | 0.047 | 0.348 |
| **ID3:ALL** | 78.51% | 0.016 | 0.046 | 0.005 | 0.406 |

# REFERENCES

[1] S. L. Prabha and A. R. M. Shanavas, "Educational Data Mining Applications," Operations Research and Applications: An International Journal, vol. 1, no. 1, pp. 23–29, 2014.

[2] C. Romero and S. Ventura, "Data mining in education," Wiley Interdisciplinary Reviews: Data Mining and Knowledge Discovery, vol. 3, no. 1, pp. 12–27, 2013.

[3] J. Jacob, K. Jha, P. Kotak, and S. Puthran, "Educational Data Mining techniques and their applications," in 2015 International Conference on Green Computing and Internet of Things (ICGCIoT), 2015, pp. 1344–1348.

[4] A. Merceron, "Educational Data Mining / Learning Analytics : Methods , Tasks and Current Trends," Proceedings of DeLFI Workshops, no. DeLFI, pp. 101–109, 2015.

[5] C. Romero, S. Ventura, M. Pechenizkiy, and R. S. J. d. Baker, Handbook of Educational Data Mining, 1st ed. CRC Press, 2011.

[6] C. Romero and S. Ventura, "Educational data mining: A review of the state of the art," IEEE Transactions on Systems, Man and Cybernetics Part C: Applications and Reviews, vol. 40, no. 6. pp. 601–618, 2010.

[7] J. Sabourin, J. Sabourin, L. Kosturko, C. Fitzgerald, and S. Mcquiggan, "Student Privacy and Educational Data Mining : Perspectives from Industry," Proceedings of the 8th International Conference on Educational Data Mining, pp. 164–170, 2015.

[8] A. U. Dr. P. Nithya, B. Umamaheswari, "A Survey on Educational Data Mining in Field of Education," International Journal of Advanced Research in Computer Engineering & Technology (IJARCET), vol. 5, no. 1, pp. 69–78, 2016.

[9] K. Sukhija, M. Jindal, and N. Aggarwal, "Educational data mining towards knowledge engineering: a review state," International Journal of Management in Education, vol. 10, no. 1, p. 65, 2016.

[10] C. Romero, J. R. Romero, and S. Ventura, "A survey on pre-processing educational data," Studies in Computational Intelligence, vol. 524. pp. 29–64, 2014.

[11] R. S. J. d. Baker, "Educational data mining: An advance for intelligent systems in education," IEEE Intelligent Systems, vol. 29, no. 3, pp. 78–82, 2014.

[12] R. Shaun, J. De Baker, and P. S. Inventado, "Chapter 4: Educational Data Mining and Learning Analytics," Springer, vol. Chapter 4, pp. 61–75, 2014.

[13] M. Berland, R. S. J. d. Baker, and P. Blikstein, "Educational data mining and learning analytics: Applications to constructionist research," Technology, Knowledge and Learning, vol. 19, no. 1–2, pp. 205–220, 2014.

[14] R. S. J. d. Baker and K. Yacef, "The State of Educational Data Mining in 2009 : A Review and Future Visions," Journal of Educational Data Mining, vol. 1, no. 1, pp. 3–16, 2009.

[15] C. Romero, S. Ventura, P. G. Espejo, and C. Hervás, "Data mining algorithms to classify students," Educational Data Mining 2008, pp. 8–17, 2008.

[16] W. Punlumjeak and N. Rachburee, "A comparative study of feature selection techniques for classify student performance," in 2015 7th International Conference on Information Technology and Electrical Engineering (ICITEE), 2015, pp. 425–429.

[17] A. A. Aziz, N. Hafieza, and I. Ahmad, "First Semester Computer Science Students ' Academic Performances Analysis by Using Data Mining Classification Algorithms," Proceeding of the International Conference on Artificial Intelligence and Computer Science(AICS 2014), pp. 100–109, 2014.

[18] L. Sibanda, C. G. Iwu, and O. H. Benedict, "Factors Influencing Academic Performance of University Students," Demography and Social Economy, vol. 24, no. 2, pp. 103–115, 2015.

[19] L. P. Steenkamp, R. S. J. d. Baker, and B. L. Frick, "Factors influencing success in first-year accounting at a South African university : A comparison between lecturers' assumptions and students' perceptions," SA Journal of Accounting Research, vol. 23, no. 1, pp. 113–140, Jan. 2009.

[20] K. Adhatrao, A. Gaykar, A. Dhawan, R. Jha, and V. Honrao, "Predicting students' performance using ID3 and C4.5 classification algorithms," International Journal of Data Mining and Knowledge Management Process, vol. 3, no. 5, pp. 39–52, 2013.

[21 ]M. Durairaj and C. Vijitha, "Educational Data mining for Prediction of Student Performance Using Clustering Algorithms," International Journal of Computer Science and Information Technologies (IJCSIT), vol. 5, no. 4, pp. 5987–5991, 2014.

[22] C. C. Aggarwal, Data classification : algorithms and applications. 2015.

[23] P. M. Goncalves, R. S. M. Barros, and D. C. L. Vieira, "On the Use of Data Mining Tools for Data Preparation in Classification Problems," in 2012 IEEE/ACIS 11th International Conference on Computer and Information Science, 2012, pp. 173–178.

[24] P. Mitra, S. Member, C. A. Murthy, and S. K. Pal, "Unsupervised feature selection using feature similarity," vol. 24, no. 3, pp. 301–312, 2002.

[25] E. Młynarska, D. Greene, and P. Cunningham, "Indicators of Good Student Performance in Moodle Activity Data," CoRR abs/1601.02975 (2016), pp. 4–7, 2016.

**Authors**

**Eng. Eid Aldikanij**, B.Sc. in Informatics Engineering [Software Engineering and Information Systems Dept.], Master Web Science (MWS) student at Syrian Virtual University, Supervisor at C ombating Money Laundering and Terrorism Financing Commission, Syria.

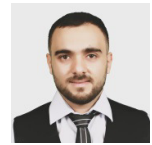

**Dr. Eng. Khalil Ajami**, (PhD) In Computer Science from Université Pierre et Marie Cu rie (Paris VI), Syrian Virtual University Vice President for Students & Admini strative Affaires, BIT Program Director.

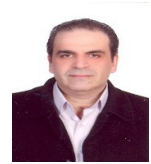